\documentclass[twocolumn,superscriptaddress,secnumarabic,amssymb, nobibnotes, aps, prb]{revtex4-2}

\usepackage{amsmath}    % need for subequations
\usepackage{graphicx}    % need for figures
\usepackage{verbatim}    % useful for program listings
\usepackage{color}           % use if color is used in text
\usepackage{subfigure}   % use for side-by-side figures
\usepackage{hyperref}    % use for hypertext links, including those to external documents and URLs
\usepackage{braket}
\usepackage{physics}
\usepackage{xfrac}
\usepackage{xspace}
\newcommand{\ruo}{RuO$_2$\xspace}
\newcommand{\bk}{\mathbf{k}}

\newcommand{\bh}{\mathbf{h}}
\newcommand{\bbm}{\mathbf{m}}

\newcommand{\bL}{\mathbf{L}}

\def\bL{\mathbf{L}}
\def\hp{\hat{p}}
\def\hd{\hat{d}}

\def\hT{\hat{T}}

\begin{document}
\title{X-ray Magnetic Circular Dichroism in \ruo}

\author{A.~Hariki}
\affiliation{Department of Physics and Electronics, Graduate School of Engineering,
Osaka Metropolitan University, 1-1 Gakuen-cho, Nakaku, Sakai, Osaka 599-8531, Japan}
\author{Y.~Takahashi}
\affiliation{Department of Physics and Electronics, Graduate School of Engineering,
Osaka Metropolitan University, 1-1 Gakuen-cho, Nakaku, Sakai, Osaka 599-8531, Japan}
\author{J.~Kune\v{s}}
\affiliation{Institute for Solid State Physics, TU Wien, 1040 Vienna, Austria}
\affiliation{Department of Condensed Matter Physics, Faculty of
  Science, Masaryk University, Kotl\'a\v{r}sk\'a 2, 611 37 Brno,
  Czechia}

\begin{abstract}
We present numerical simulation of the X-ray magnetic circular dichroism (XMCD) 
of the $L_{2,3}$ and $M_{2,3}$ edges of Ru in antiferromagnetic \ruo using
a combination of density functional + dynamical mean-field theory and configuration interaction treatment of Anderson impurity model. We study the dependence of the dichroic spectra on the orientation of the N\'eel vector and
discuss it in the context of altermagnetism. Approximate equivalence between
the XMCD spectra for geometries with X-rays propagating parallel and perpendicular to the N\'eel vector is found and shown to be exact in absence 
of valence spin-orbit coupling and core-valence multipolar interaction.

\end{abstract}

\maketitle

\section{Introduction}
A metallic compensated magnet \ruo has attracted a considerable attention recently. The combination of its rutile structure
with an antiparallel ordering of spin moments gives rise to a number of phenomena such as the anomalous Hall effect (AHE) \cite{Smejkal22b,Smejkal20,Samanta20,Naka20,Hayami21,Mazin21,Betancourt23,Naka22} or charge-spin conversion effects \cite{Smejkal22a,Naka19,Gonzalez-Hernandez21,Naka21,Ma21,Smejkal22c,Smejkal22,Lovesey2022}, and strongly spin-polarized electronic band structures \cite{Ahn19,Hayami19,Smejkal22a,Smejkal20,Yuan20,Yuan21,Hayami20,Smejkal22,Mazin21,Liu22,Jian23}, unusual among compensated collinear magnets.
Moreover, some of these effects depend on the orientation of the N\'eel vector $\bL$ and may be switched by manipulating it.
The name altermagnet \cite{Smejkal22a,Smejkal22} was introduced for \ruo and similar materials
to reflect the alternating spin polarization in the momentum space. 

Altermagnetism is a non-relativistic concept, i.e., it relies on the spin rotation $SU(2)$ symmetry
in the absence of the spin-orbit coupling (SOC). However, the above transport 
and charge-excitation effects are possible only if altermagnetic order and SOC are present simultaneously. The SOC thus plays a dual role. On one hand, it allows the altermagnetic order to 
be detectable through some of the technologically interesting transport effects. On the other hand,
it gives rise to magneto-crystalline anisotropy and possibly a weak ferromagnetism
with a small net magnetization $\bbm$. To prove the altermagnetic origin of transport phenomena, 
e.g., AHE, one must not only measure a finite signal, but also distinguish the altermagnetic
contribution associated with $\bL$ from the contribution due to weak ferromagnetism or possibly the external magnetic field associated with $\bbm$, which requires a careful quantitative analysis~\cite{Feng2022}. 

%A net magnetization $\bbm$ also arises if an external field is used to overcome the
%magneto-crystalline anisotropy in order 
%to explore the orientation dependence of $\bL$. 
%As a result the anomalous Hall coefficient or magneto-optical effects generally have two contributions.
%One, depending on $\bL$, is specific to altermagnets. The second one, proportional to $\bbm$, 
%induced by external field or due to weak ferromagnetism, 
%is present in any material with finite net magnetization.
%Since both contributions vanish without SOC, disentangling the altermagnetic 
%part is a non-trivial quantitative question~\cite{Feng2022}.

%%%%%%%%%%%%%%%%%%%%%%%%%%%%%%%%%%%%
%%%%%%%%%%%%%%%%%%%%%%%%%%%%%%%%%%%%
\begin{figure}[t]
\includegraphics[width=0.93\columnwidth]{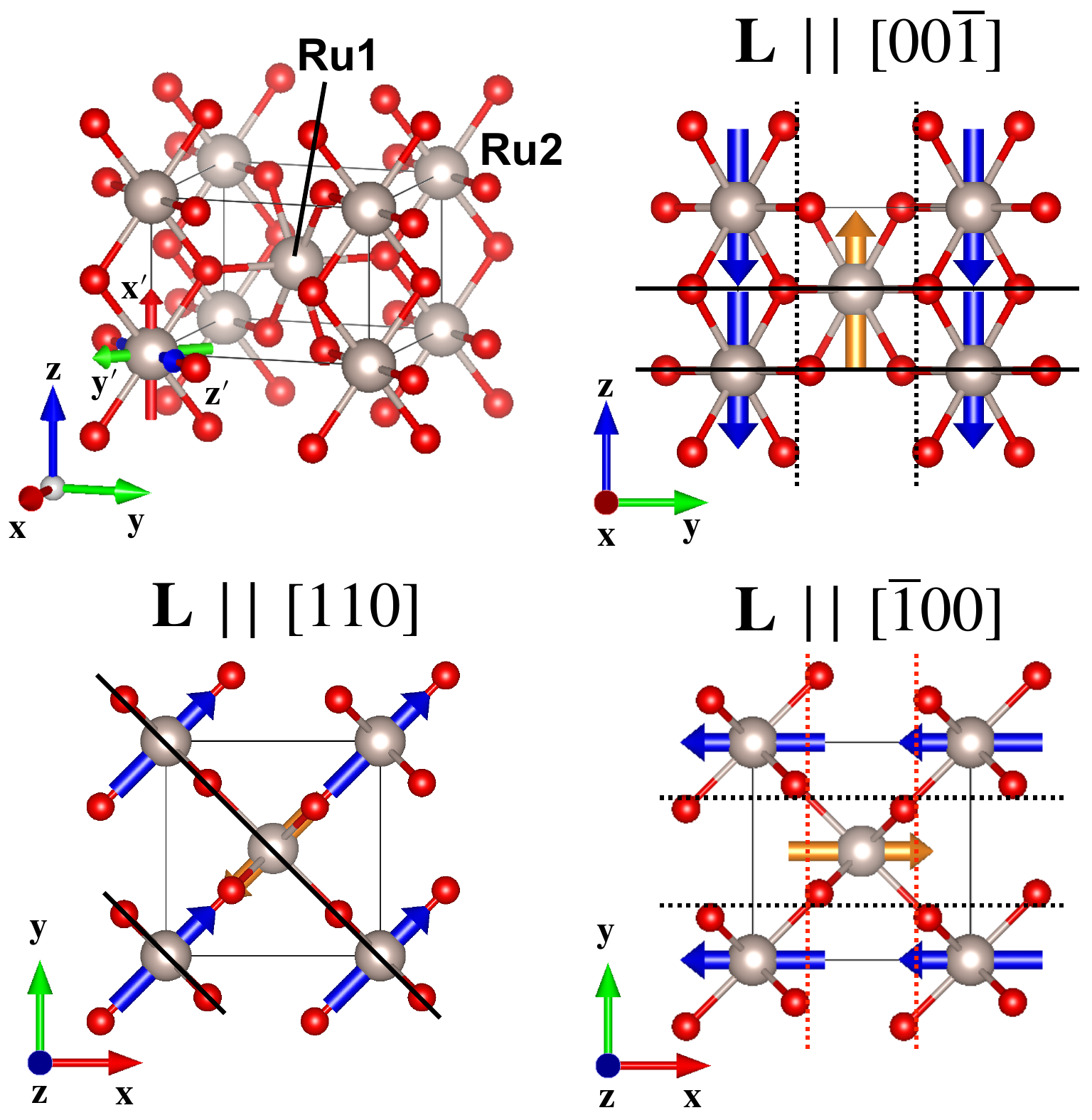}
    \caption{Crystal structure of \ruo and orientations of magnetic structures studied
    in this work. Solid (dashed) lines indicate the mirror (glide) plane $m$.}
\label{fig_struct}
\end{figure}
%%%%%%%%%%%%%%%%%%%%%%%%%%%%%%%%%%%%
%%%%%%%%%%%%%%%%%%%%%%%%%%%%%%%%%%%%

%%%%%%%%%%%%%%%%%%%%%%%%%%%%%%%%%%%%
%%%%%%%%%%%%%%%%%%%%%%%%%%%%%%%%%%%%
\begin{figure*}
\includegraphics[width=1.99\columnwidth]{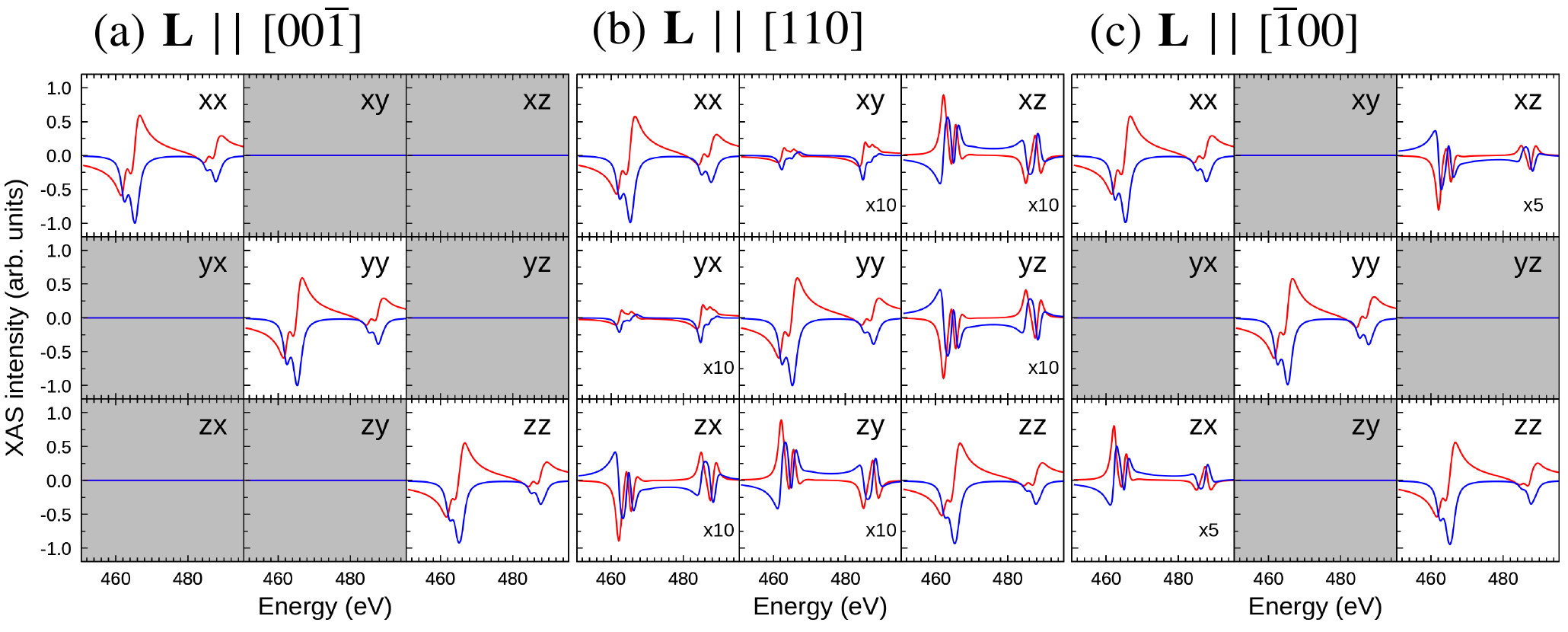}
    \caption{Real (red) and imaginary (blue) part of the optical conductivity tensor at Ru $M_{2,3}$ edge for different orientations of the N\'eel vector $\bL$. Here, the Ru$^{4+}$ atomic model ($J=0.50$~eV) is used.}
\label{fig_abc}
\end{figure*}
%%%%%%%%%%%%%%%%%%%%%%%%%%%%%%%%%%%%
%%%%%%%%%%%%%%%%%%%%%%%%%%%%%%%%%%%%

The X-ray magnetic circular dichroism (XMCD) provides an alternative. 
The XMCD is an odd magneto-optical effect, which arises from the same anti-symmetric part of the conductivity tensor ${\boldsymbol{\sigma}^a=\tfrac{1}{2}(\boldsymbol{\sigma}-\boldsymbol{\sigma}^T)}$ as the valence band phenomena, such as AHE or magneto-optical effects in the visible range and therefore follows the same symmetry rules.
However, the XMCD, especially in lighter elements, arises due to the SOC in the core state and is only weakly affected by the valence SOC reflecting the typically small orbital moments~\cite{Thole1992,Weller1995,Kunes2003}. 
The theoretical altermagnetic contribution to XMCD can be obtained as the XMCD with the valence SOC switched off. In other words, in the X-ray spectral range it is the core SOC which allows the altermagnetic order to be expressed in the magneto-optical spectra, while the valence SOC, which causes the spurious weak ferromagnetism, has only a marginal contribution~\cite{Hariki2023}. In the optical or transport effects
the valence SOC plays a dual role, which cannot be resolved by simply switching it off.

In this article we compute XMCD at $L_{2,3}$ and $M_{2,3}$ edges of Ru
for various orientations of the N\'eel vector $\bL$. %These correspond to excitation of a $2p$ and $3p$ hole, respectively. Since both involve electron transfer to the
%$4d$ states, the only difference between these excitations is the strength of the core--valence
%interaction and the size of the core SOC~\footnote{The excitations take place as vastly different
%energies and have different cross sections, none of which affects the shape of the spectra.}.
Recently, Sasabe {\it et al.}~\cite{Sasabe23} calculated XMCD in \ruo 
using the atomic model in an external Zeeman field, and discussed their results in the context of strong-coupling (insulator) picture of excitonic magnet~\cite{Khaliullin2013}. Our calculation of XMCD
is performed for the antiferromagnetic metallic state~\cite{Ahn19} obtained with the density functional plus dynamical mean-field theory (DFT+DMFT)~\cite{Metzner1989,Georges96,Kotliar06}. The method identifies a Fermi surface instability as the origin of the antiferromagnetism of metallic RuO$_2$~\cite{Ahn19}.

%Therefore the spectra at $M$- and $L$-edges have similar, but not identical shapes.

\section{Computational Method}
Starting with a density functional theory (DFT) calculation for the experimental structure of RuO$_2$~\cite{Berlijn17} using Wien2k~\cite{wien2k}, we construct a multi-band Hubbard model~\cite{wien2wannier,wannier90} spanning the Ru $4d$ states. 
The Ru 4$d$ crystal field levels %are split by the crystal field 
%by surrounding ligands as well as anisotropy in the hybridization 
are summarized in Table~\ref{tab_ene}. 
%The XMCD line shape is sensitive to the 4$d$ level splitting.
The electron-electron interaction within the Ru $d$-shell is parametrized by $U=$~3.0~eV and $J=$~0.45~eV. % in the main results.
The interaction parameters affect the magnitude of the ordered Ru  moment~\cite{Ahn19}, and thus the XMCD intensities, as examined in the Supplemental Material (SM)~\cite{sm} (see also references \cite{Matsubara00,wang09,boehnke11,hafermann12,werner06,jarrell96} therein).
We use the same DMFT implementation for the multi-band Hubbard model  %of the DMFT self-consistent calculation 
as in Refs.~\onlinecite{Hariki2017,Hariki18,Hariki20}.  
Material-specific details can be found in SM~\cite{sm}. 

The XMCD spectra are calculated for i) Ru$^{4+}$ atomic model and ii) DMFT + Anderson impurity model (AIM) using the method of Refs.~\onlinecite{Hariki2017,Hariki18,Hariki20} based on a configuration interaction impurity solver.
The two models implement the same atomic Hamiltonian, which is coupled to an electronic bath via the hybridization function $\Delta(\omega)$ in (ii). 
The ordered Ru moments are generated by the (self-consistently determined) spin-polarized bath
in (ii). In (i) we impose a Zeeman field chosen to generate the same moment as obtained in (ii).
Detailed comparison of the two models in SM~\cite{sm} reveals similar spectra.
The computationally cheap atomic model is therefore used for the symmetry analysis of the results.

The SOC 
%within the Ru 4$d$ shell 
is not included in the DMFT self-consistent calculations. 
%thus its effect is not present in the 
%hybridization function $\Delta(\omega)$ (Weiss-field) and 
%the magnetocrystalline anisotropy is not captured by the simulations. 
However, SOC within the Ru 4$d$ shell is included in the AIM 
%as well as atomic model 
when computing the XMCD intensities. The spin-polarized hybridization densities $\Delta(\omega)$ are transformed to capture the desired N\'eel vector orientations in these simulatons.

%%%%%%%%%%%%%%%%%%%%%%%%%%%%%%%%%%%%%%%%%%%%%%%%%
      \begin{table}[h]
            \centering  
            \begin{tabular}{c | c c c c c c }
            \hline\hline
            &{\ }$d_{xy}${\ }   & {\ }$d_{3z^2-r^2}${\ }  & {\ }$d_{x^2-y^2}${\ }  & {\ }$d_{zx}${\ }  & {\ }$d_{yz}${\ }
            \\[+3pt]
            \hline\\[-10pt]
            {\quad}Energy (eV){\ }{\quad} &{\ }3.283{\ }&{\ }3.522{\ }& -0.127{\ }& 0.036& {\ }0.067{\ }\\[+2pt]
            \hline\hline
            \end{tabular}
            \vspace*{+0.2cm}
            \caption{The Ru 4$d$ orbital energies in RuO$_2$ derived from the DFT calculation for the experimental rutile structure. The 4$d$ orbitals are represented in a local coordinate ($x'y'z'$) in the left top panel of Fig.~\ref{fig_struct}.}
            \label{tab_ene}
        \end{table}        
%%%%%%%%%%%%%%%%%%%%%%%%%%%%%%%%%%%%%%%%%%%%%%%%%

%\textcolor{red}{Maybe we shall first introduce Fig.~\ref{fig_mcd}, pointing out the presence of a shoulder feature corresponding to an excitation into a $t_{2g}$ state in both theory and experiment, with an emphasis that the XMCD is found on that low-energy feature.}

%DMFT with these parameters reproduces well the valence-band both in experiment~\cite{Sato99} and theory~\cite{pfjr:23} as well as Mn $2p$ core-level x-ray photoemission spectra~\cite{Jian10}, see SM~\cite{sm}.

\section{Results}
The XMCD is the difference of
the absorption spectra for the right-hand and left-hand
circularly polarized light propagating along the direction $\hat{\bk}$. It is convenient to view the antisymmetric part of the conductivity tensor $\boldsymbol{\sigma}$ as an axial (Hall) vector ${\bh(\omega) = (\sigma^a_{zy}(\omega), \sigma^a_{xz}(\omega), \sigma^a_{yx}(\omega))}$. For simplicity we will not show
the $\omega$-dependence from now on, but indicate the dependence on $\hat{\bk}$ as well as the orintation
of the N\'eel vector $\bL$
\begin{equation}
    \Delta F(\hat{\bk},\bL)=F^+(\hat{\bk},\bL)-F^-(\hat{\bk},\bL)=
    2\operatorname{Im}\bh(\bL)\cdot \hat{\bk}.
\end{equation}
The Hall vector $\bh(\bL)$ contains information about XMCD for any direction $\bk$, but depends on the crystallographic orientation
of $\bL$. In Fig.~\ref{fig_abc} we show the elements of $\boldsymbol{\sigma}$ at the Ru $M_{2,3}$ edge
for $\bL$ along $[00\bar{1}]$, $[110]$ and $[\bar{1}00]$ directions.

Thanks to the local nature of core-level excitations the observed signal is given by a sum of 
contributions from the two Ru sites~\cite{Winder20}. 
%In the non-relativistic case (no valence SOC),
% the local moments are perfectly anti-parallel and
%there is no energetically preferred orientation of $\bL$
%(no magneto-crystalline anisotropy). 
%Nevertheless, XMCD may arise thanks to the core-level SOC and possibly core-valence exchange interaction~\cite{Hariki2023}.
Depending on the orientation of $\bL$, 
%when relativistic symmetry
%, i.e, locking spin and orbital symmetries together, 
%is taken into account 
the two Ru sites are connected by some relativistic symmetry operation
or not.
%are not symmetry equivalent or ii) the
%two Ru sites are coupled by a symmetry operation. 
In the former case XMCD may vanish due to cancellation between the site contributions even if these are non-zero.
In the latter case a non-zero XMCD exists if allowed locally by the site symmetry.
%In case (ii) XMCD may vanish due to cancellation between the site contributions even if these are non-zero.

\subsection{Symmetry considerations}
%It is convenient to view the antisymmetric part of $\boldsymbol{\sigma}$ as a axial (Hall) vector ${\bh(\omega) = (\sigma^a_{zy}(\omega), \sigma^a_{xz}(\omega), \sigma^a_{yx}(\omega))}$. The XMCD for x-rays with wavevector $\bk$
%is then given as $\tfrac{2\operatorname{Im}\bh(\omega)\cdot \bk}{|k|}$.
%\textcolor{red}{ATSUSHI, CAN YOU CHECK WHETHER THE SIGN IS CONSISTENT WITH YOU CONVENTION/FIGS!}
%\textcolor{blue}{$\rightarrow$ I WILL DO THIS WEEK SOON.}
In the following we will make use of mirror symmetries to discuss the  dependence of the XMCD signal, i.e.~$\bh(\bL)$ as a function of $\bL$.
The mirror (glide) planes are marked in Fig.~\ref{fig_struct}. 
XMCD is forbidden if there is a mirror plane parallel to x-ray wave vector $\bk$, since it maps the
right-hand circular polarization on the left-hand one and vice versa.
Since the local moments transform as axial vectors,
a crystallographic mirror (glide) plane $m$ is retained as an element of the relativistic symmetry group
either if $m$ is parallel to $\bL$ and maps the magnetic sublattices on one another or
$m$ is perpendicular to $\bL$ and maps each magnetic sublattice on itself. 
Using these simple rules we can understand behavior of the calculated XMCD.

%%%%%%%%%%%%%%%%%%%%%%%%%%%%%%%%%%%%
%%%%%%%%%%%%%%%%%%%%%%%%%%%%%%%%%%%%
\begin{figure}[t]
\includegraphics[width=0.98\columnwidth]{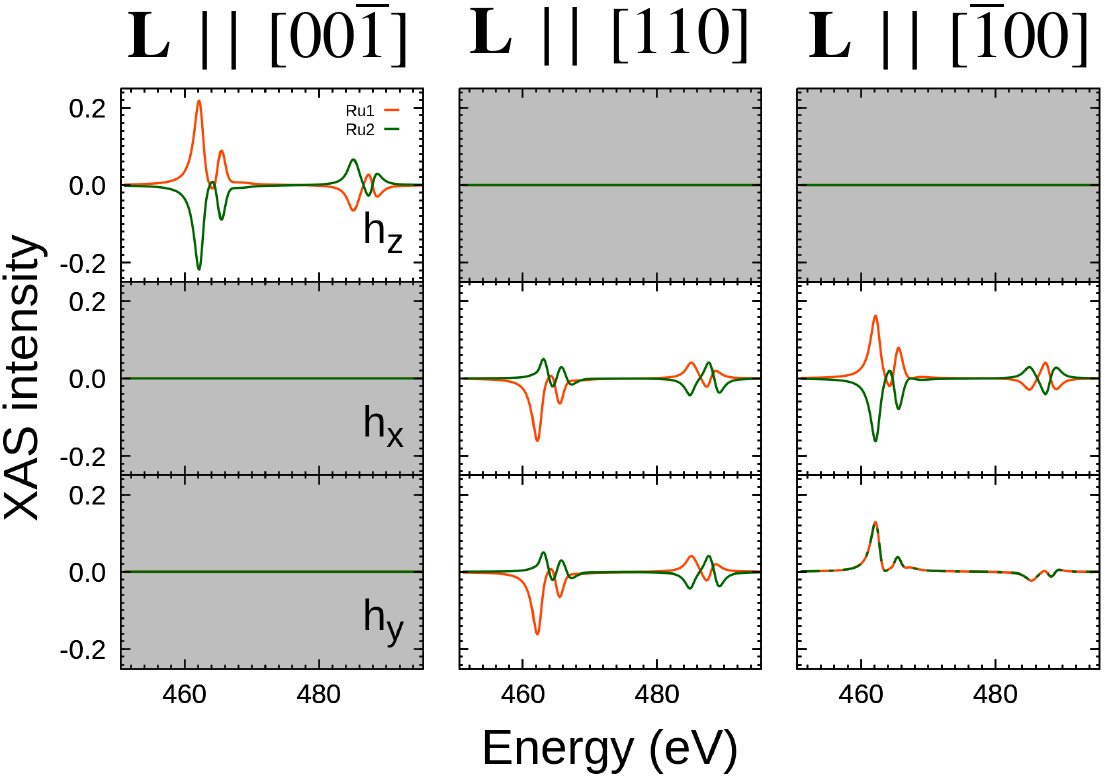}
    \caption{The imaginary part of the Hall vectors $\bh^{(i)}$ at the Ru $M_{2,3}$-edge for site 1 (yellow) and site 2 (green) obtained for different orientations of the N\'eel vector $\bL$. Here, the Ru$^{4+}$ atomic model ($J=0.50$~eV) is used. The XMCD spectra of the Ru$^{4+}$ atomic model are found in SM~\cite{sm}.}
\label{fig_site}
\end{figure}
%%%%%%%%%%%%%%%%%%%%%%%%%%%%%%%%%%%%
%%%%%%%%%%%%%%%%%%%%%%%%%%%%%%%%%%%%

For $\bL\parallel [110]$ the two Ru sites are inequivalent. There is no symmetry relationship between the XMCD 
contributions from the two Ru sites and therefore no cancellation between the site contributions $\bh^{(i=1,2)}$ as shown in Fig.~\ref{fig_site}.
The total XMCD signal is non-zero unless XMCD is symmetry-forbidden locally on each site. 
The mirror plane $m_{(110)}$ maps each Ru site on itself and is perpendicular to the
the local moments, which implies that only $\bh^{(i)}\parallel [110]$ is allowed.
%This is encoded in  $\sigma^{(i)}_{ab}=\sigma^{(i)}_{ba}$ and $\sigma^{(i)}_{ac}=-\sigma^{(i)}_{ca}=-\sigma^{(i)}_{bc}=\sigma^{(i)}_{cb}$.

For both $\bL\parallel [00\bar{1}]$ and $\bL\parallel [\bar{1}00]$ the two Ru sites are equivalent. For $\bL\parallel [00\bar{1}]$
the presence of two non-parallel mirror planes, the glide planes $n_{(100)}$ and $n_{(010)}$
mapping the two Ru sites on each other, forbids the total XMCD signal for any direction
of $\bk$: $\bh^{(1)}=-\bh^{(2)}$.
Moreover, the mirror plane $m_{(001)}$ implies that $h^{(i)}_x=h^{(i)}_y=0$ on each site as shown in Fig.~\ref{fig_site}.
For $\bL\parallel [\overline{1}00]$, the glide plane $n_{(010)}$ implies
$h^{(1)}_z=-h^{(2)}_z$ and $h^{(1)}_x=-h^{(2)}_x$, see Fig.~\ref{fig_site}.
The glide plane $n_{(100)}$, indicated by red dashed lines in Fig.~\ref{fig_struct}, which maps the two Ru sites on each other, must be coupled with the time reversal
$n_{(100)}\mathcal{T}$ in order to be a symmetry operation. Therefore the two Ru sites
yield the same contribution  
$h^{(1)}_y=h^{(2)}_y$, see Fig.~\ref{fig_site}. Finally, we observe that $h^{(i)}_z=0$ locally thanks to the mirror
plane times time-reversal operations $m_{(001)}\mathcal{T}$.

%%%%%%%%%%%%%%%%%%%%%%%%%%%%%%%%%%%%
%%%%%%%%%%%%%%%%%%%%%%%%%%%%%%%%%%%%
\begin{figure}
\includegraphics[width=0.95\columnwidth]{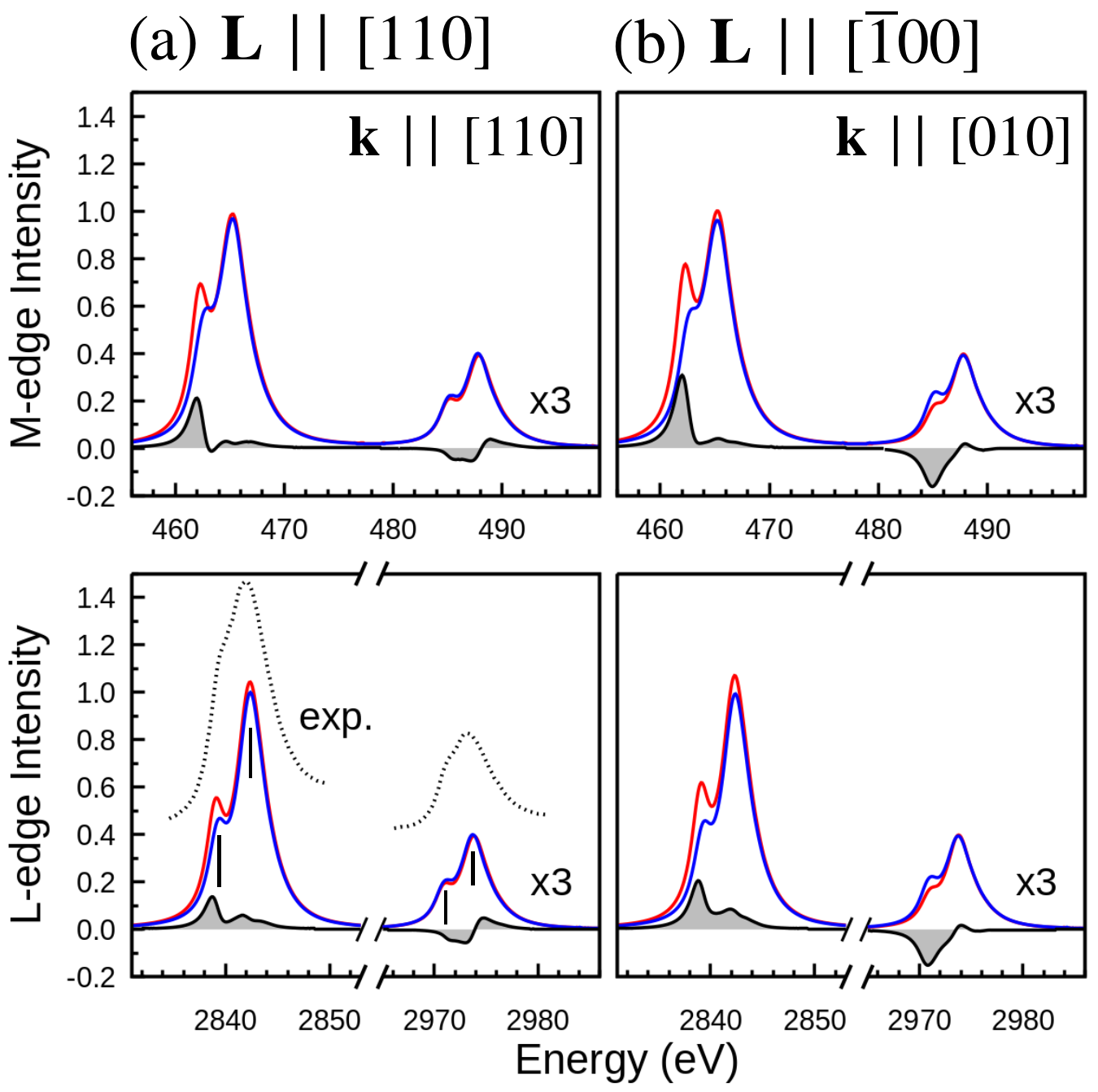}
    \caption{
    The XAS calculated for the two circular polarizations (red and blue) at the Ru $M_{2,3}$-edge (top) and $L_{2,3}$-edge (bottom) together with the XMCD intensities (shaded) calculated for different orientations of the N\'eel vector $\bL$ and x-ray propagation vector $\bf k$ using the LDA+DMFT AIM method. The calculated spectral intensities are broadened by Lorentizan of 1.0~eV (HWHM).
    The experimental $L_3$-edge x-ray absorption spectrum is taken from Ref.~\onlinecite{Hu00}. The $L_3$ main line and shoulder feature are indicated by vertical lines.
%    The XMCD intensities at the $M_2$- and $L_2$-edge are magnified by a factor of 3.
    }
\label{fig_mcd}
\end{figure}
%%%%%%%%%%%%%%%%%%%%%%%%%%%%%%%%%%%%
%%%%%%%%%%%%%%%%%%%%%%%%%%%%%%%%%%%%

\subsection{$L$- and $M$-edge of \ruo}
Fig.~\ref{fig_mcd} shows Ru $L_{2,3}$- and $M_{2,3}$-edge absorption spectra  with left and right circularly-polarized x-rays and the XMCD intensities calculated using the LDA+DMFT AIM method. The $L$- and $M$-edge spectra have similar shapes consisting of a main-peak and a shoulder, indicated by vertical lines~\footnote{The similarity of the $L$- and $M$-edge is not surprising since they both correspond to $p$--$d$ excitations, which are different only by core-valence interaction strength.},
and agree well with the the available experimental absorption data
on the $M_{2,3}$-edges~\cite{Hu00}.
%\textcolor{blue}{\bf (moved) The calculated \textcolor{red}{\bf $M_{2,3}$-edges} exhibit a fair agreement with the available experimental absorption data~\cite{Hu00}.}
%The shoulder (main line) is primarily associated with excitations to the empty Ru $t_{2g}$ ($e_g$) states.  
The shoulders arise from the Ru $t_{2g}$ bands, which 
host the spin moment. The $e_{g}$ bands are spin-split but empty. 
Consequently, the XMCD intensities are largely concentrated in the shoulder for all $M_{2,3}$- and $L_{2,3}$-edges. 
A similar shape of the XMCD spectra was observed on the $M_{2,3}$-edge XMCD of SrRuO$_3$,  a prototypical ferromagnetic metallic Ru$^{4+}$ oxide~\cite{Okamoto07}.
%providing validation for predicting the XMCD which, to the best of our knowledge, has not been measured yet.

Our results show that  a non-zero XMCD exists for $\bL\parallel [110]$ and
any $\bk$ not perpendicular to $\bL$ as well as $\bL\parallel[\bar{1}00]$ and
$\bk$ not perpendicular to $[010]$. On the other hand, the XMCD signal vanishes for the experimental easy axis $\bL\parallel [00\bar{1}]$. It was shown that $\bL$ can be tilted in the $[110]$ using external magnetic field~\cite{Feng2022}.
While we have included the valence SOC in calculation of the XMCD spectra
using both methods (i) and (ii), we did not include the magnetocrystalline
anisotropy in our calculations. Therefore we can only speculate about the
external field effect. Looking at the site contributions for the $\bL\parallel [110]$ in Fig.~\ref{fig_site} we observe that site 1 dominates over site 2.
Assuming that the external field has only moderate influence on the size of the
local moments, we may conclude that it does not alter the present result 
substantially.

\subsection{No valence SOC, no core-valence multipole interaction}
\begin{figure}[t]
\includegraphics[width=0.95\columnwidth]{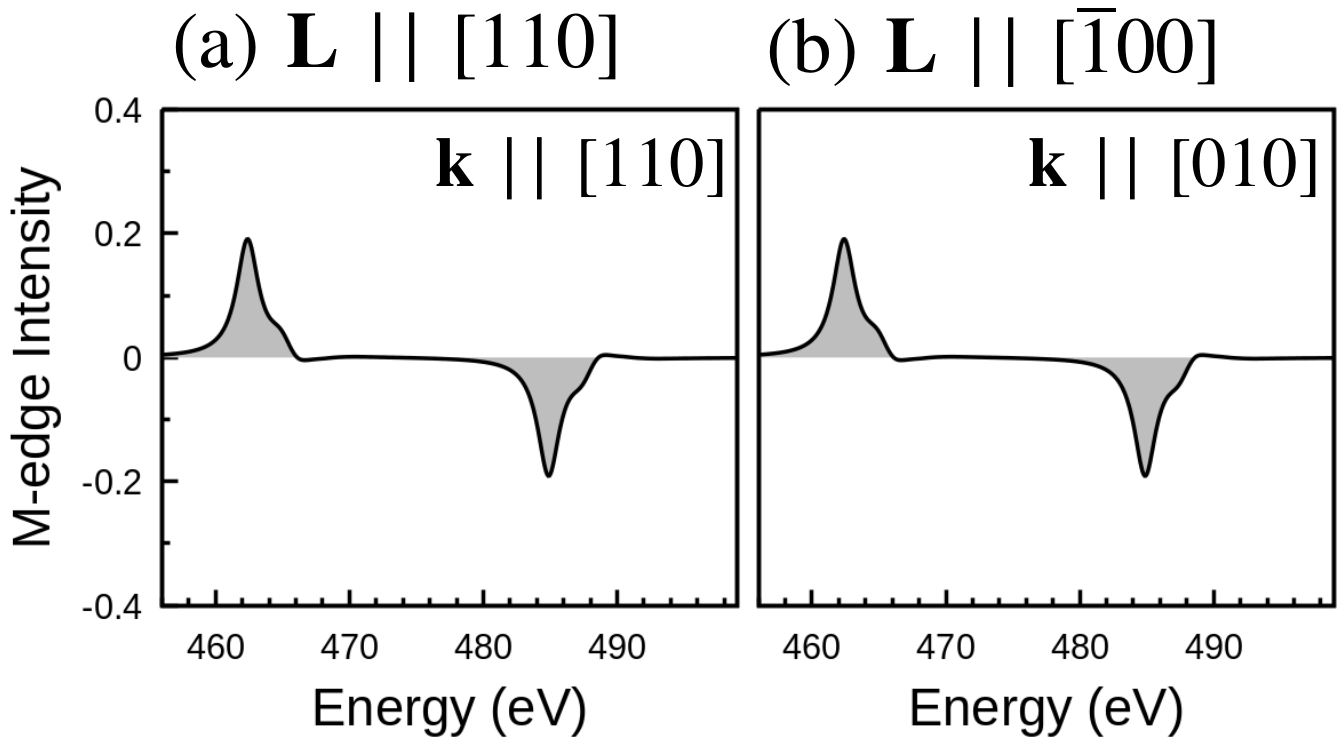}
    \caption{Ru $M_{2,3}$-edge XMCD intensities $\Delta F^{(a)}$ in case (a) and $\Delta F^{(b)}$ in case (b) calculated without SOC within the Ru 4$d$ shell and 3$p$--4$d$ core-valence multiplet interaction for the two different N\'eel vectors $\bL$ and x-ray propagation vectors $\hat{\bk}$ by the LDA+DMFT AIM method.}
\label{fig_mcd_nsoc}
\end{figure}

 Comparing the XMCD spectra in Fig.~\ref{fig_mcd}a and Fig.~\ref{fig_mcd}b we observe a similarity pronounced in particular on the $L_3$ and $M_3$ edges.
In Fig.~\ref{fig_mcd_nsoc} we show that this is not accidental. In Ref.~\onlinecite{Hariki2023} we have observed that 
turning off the valence SOC and the multipole part of the core-valence interaction may change the XMCD spectra qualitatively. This is because the Hamiltonian without these interactions possesses a higher symmetry. In the present case, switching off the valence SOC and the multipole part of the core-valence interaction results in  equality of the two spectra $\Delta F^{(a)}(\omega)=\Delta F^{(b)}(\omega)$. This may appear surprising given the apparently
different geometry in case (a) $\bk\parallel\bL$ while in case (b) $\bk\perp\bL$. 
In Appendix.~\ref{app:nsoc} we provide an analytic proof, which uses only the presence of two-fold axes along the $[110]$ and $[1\bar{1}0]$ directions. 

%%%%%%%%%%%%%%%%%%%%%%%%%%%%%%%%%%%%
%%%%%%%%%%%%%%%%%%%%%%%%%%%%%%%%%%%%
%%%%%%%%%%%%%%%%%%%%%%%%%%%%%%%%%%%%
%%%%%%%%%%%%%%%%%%%%%%%%%%%%%%%%%%%%

\section{Conclusions}
We have calculated the XMCD spectra on the Ru $L_{2,3}$ and $M_{2,3}$ edges of antiferromagnetic RuO$_2$ using the LDA+DMFT approach. The  present  XMCD spectra  differ from the recent calculations using the atomic model~\cite{Sasabe23}. The origin of this discrepancy can be traced to  different values of the crystal field and  Hund's coupling parameters. 
%In our calculations we have not found any indication of excitonic magnetism suggested in Ref.~\onlinecite{Sasabe23}. 
%\textcolor{red}{\bf Though the excitonic magnetism picture cannot be applied to RuO$_2$, our current DMFT without SOC in self-consistent calculation probably could not detect if it realizes (though it should not realize in real RuO$_2$). So, perhaps we need to be a bit careful for this strong statement for them? Maybe we mention the conceptual problem in their treatment (atomic model + simple spin Zeeman field) somewhere?}
We have analyzed the symmetry of the XMCD spectra for various orientations of the N\'eel vector. The results apply to any collinear antiferromagnet with the rutile structure. No XMCD is allowed for the easy axis $[001]$ orientation of the N\'eel vector $\bL$. We have predicted the XMCD spectra for the experimentally accessible $[110]$ orientation.

%%%%%%%%%%%%%%%%%%%%%%%%%%%%%%%%%%%%
%%%%%%%%%%%%%%%%%%%%%%%%%%%%%%%%%%%%
%\begin{figure}[t]
%\includegraphics[width=0.98\columnwidth]{RuO2_figures/fig_mcd.pdf}
%    \caption{\textcolor{red}{to be moved to SM} Ru $M_3$-edge XMCD intensities calculated for different orientation of the N\'eel vectors $\bL$ and x-ray wave vectors $\bf k$.}
%\label{fig_mcd}
%\end{figure}
%%%%%%%%%%%%%%%%%%%%%%%%%%%%%%%%%%%%
%%%%%%%%%%%%%%%%%%%%%%%%%%%%%%%%%%%%

%%%%%%%%%%%%%%%%%%%%%%%%%%%%%%%%%%%%
\begin{comment}
%%%%%%%%%%%%%%%%%%%%%%%%%%%%%%%%%%%%
\begin{figure}[t]
\includegraphics[width=0.93\columnwidth]{RuO2_figures/fig_demo.pdf}
    \caption{Ru $M_3$-edge XMCD intensities calculated without the SOC on the Ru 4$d$ valence shell for the selected  N\'eel vectors $\bL$ and x-ray wave vectors $\bf k$ by the LDA+DMFT AIM method $U=3.0$~eV and $J$=0.45~eV.}
\label{fig_demo}
\end{figure}
%%%%%%%%%%%%%%%%%%%%%%%%%%%%%%%%%%%%
%%%%%%%%%%%%%%%%%%%%%%%%%%%%%%%%%%%%
\end{comment}

\begin{acknowledgements}
We thank Andriy Smolyanyuk and Anna Kauch for discussions and critical reading of the manuscript. This work was supported by 
JSPS KAKENHI Grant Numbers 21K13884, 21H01003, 23K03324, 23H03817
(A.H.), and by the project Quantum materials for applications in sustainable technologies (QM4ST), funded as project No. CZ.02.01.01/00/22\_008/0004572 by Programme Johannes Amos Commenius, call Excellent Research (J.K.).
\end{acknowledgements}

\appendix
\section{XMCD with no core-valence exchange and no valence SOC}
\label{app:nsoc}
Here we prove analytically the numerical results of Fig.~\ref{fig_mcd}. We start with several definitions. The circular dichroism ${\Delta F(\hat{\bk},\bL)=
F^+(\hat{\bk},\bL)-F^-(\hat{\bk},\bL)}$ is the difference of the absorption spectra for the 
right-hand and left-hand circularly polarized light can be obtained from 
the Fermi golden rule
\begin{equation}
\label{eq:fsum}
F^\pm(\hat{\bk},\bL)=\sum_{f} \left|\expval{f_\bL|\hT^{\pm}_{\hat{\bk}}|i_\bL}\right|^2\!\delta\left(\omega\!-\!E_{fi;\bL}\right).
\end{equation}
Here $|i_\bL\rangle$ and $|f_\bL\rangle$ are the eigenstates of the Hamiltonian, $E_{fi;\bL}$ is the excitation energy, and $\hT^{\pm}_{\hat{\bk}}$ are the dipole operators for the right- and left-hand polarization
with respect to propagation vector $\hat{\bk}$.
Thanks to the immobility of the core hole
the x-ray absorption spectrum is an sum over site contributions.

\begin{figure}
    \centering
    \includegraphics[width=0.5\linewidth]{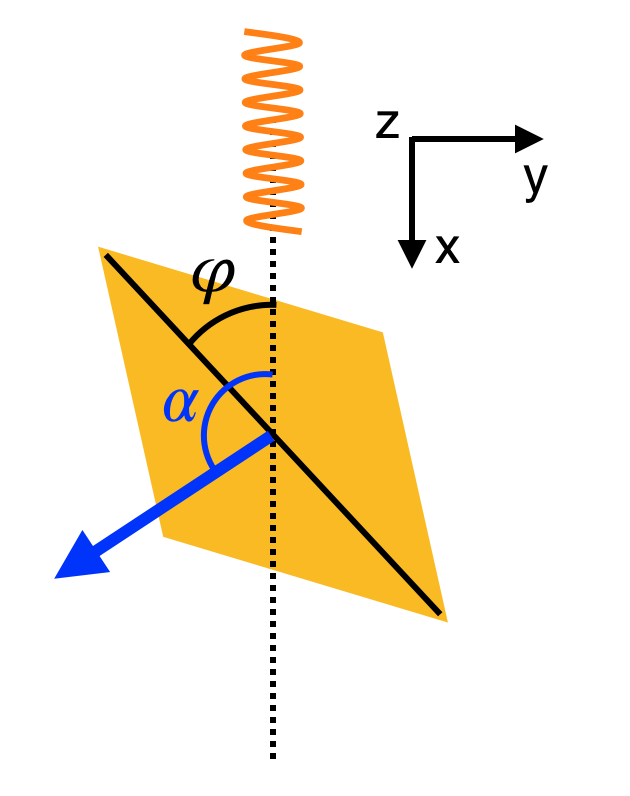}
    \caption{Geometry used to prove the equality of $\Delta F^{(a)}$ and $\Delta F^{(b)}$ numerically observed in Fig.~\ref{fig_mcd}. The Ru atom is located in the center, the parallelogram (long axis parallel to $[110]$) represents the orientation of the crystal.}
    \label{fig:cartoon}
\end{figure}

We will use the geometry of Fig.~\ref{fig:cartoon} with the x-rays coming along the $x$-axis
and the quantization $z$-axis of spin and angular momenta parallel to the
crystallographic $c$-axes. To evaluate XMCD $\Delta F^{(a)}$ and $\Delta F^{(b)}$ for the geometries for Fig.~\ref{fig_mcd}a and Fig.~\ref{fig_mcd}b we will
vary the angles $\varphi$ (orientation of the crystal) and $\alpha$ (orientation of the local moment)
\begin{equation}
\label{eq:comparsion}
\begin{split}
        \Delta F^{(a)}&=\frac{1}{2}\left(\Delta F(0,0)+\Delta F(\tfrac{\pi}{2},\pi)\right)\\
        \Delta F^{(b)}&=\Delta F(\tfrac{\pi}{4},\tfrac{\pi}{2}).
    \end{split}
\end{equation}
Here we use the angles $\varphi$ and $\alpha$ in $\Delta F(\varphi,\alpha)$ to represent 
$\hat{\bk}$ and $\bL$ in the geometry of Fig.~\ref{fig:cartoon}.
Using the relations between the dipole operators $\hT^x$, $\hT^y$ and $\hT^z$ 
    \begin{equation*}
    \begin{split}
        T^+&=%\tfrac{1}{\sqrt{2}}\left(
        -T^x-iT^y
        %\right)
        \\
        T^-&=
        %\tfrac{1}{\sqrt{2}}\left(
        -T^x+iT^y
        %\right)
        \\
        T^0&=T^z,
    \end{split}
    \end{equation*}
we express the dipole operators $T^\pm_{\hat{x}}$ for the circularly polarizated light propagating along the $x$-axis,
helicity basis for $\hat{\bk}\parallel x$,  using dipole operators 
in the helicity basis for $\hat{\bk}\parallel z$\footnote{We drop the $\hat{z}$ subscript in $T^\pm_{\hat{z}}$ for sake of readability.}
\begin{equation*}
    T^\pm_{\hat{x}}=-T^y\mp iT^z=i\left(\frac{T^+-T^-}{2}\pm T^0\right).
\end{equation*}
The expression (\ref{eq:fsum}) for XMCD then takes the form  
    \begin{equation}
    \label{eq:A}
    \begin{split}
       \Delta F(\varphi,\alpha)=&\sum_{f} \expval{f_{\varphi,\alpha}|\hT^{+}-\hT^{-}|i_{\varphi,\alpha}}\expval{i_{\varphi,\alpha}|\hT^0|f_{\varphi,\alpha}}\\
       &\times
       \delta\left(\omega\!-\!E_{fi}\right)+c.c.   \\
       \equiv & (T^+-T^-)\overline{T^0}+c.c.
           \end{split}
    \end{equation}
    Here, the polarization of the dipole operators is taken with respect to the $z$-axis. We will use
    the shorthand notation for the matrix elements of the dipole operators shown on the third line from now on.
    In absence of the valence SOC  and core-valence multipole interaction, the excitation energy
    does not depend on the orientation of $\bL$, i.e., the angle $\alpha$ and the eigenstates
    for arbitrary angles $\varphi$ and $\alpha$ can be obtained
    %In the following we will use the shorthand expression on the second line since we will be concerned only with the transformation properties of the matrix elements. Note that the different form of
%(\ref{eq:fsum}) and (\ref{eq:A}) results from light coming along the $x$-axis, but $\hT^pm$ taken with respect to the $z$-axis. The purpose of this apparently cumbersome geometry is to perform rotations in the $xy$-plane easily.
from those for $\varphi=0$ and $\alpha=0$ by 
%(Weiss field) can be transformed into each other by a rotation of valence spin. 
%In the following we will use 
a joint $z$-axis rotation $\mathcal{C}(\varphi,\alpha)$ of core spin, core orbitals and valence orbitals by angle $\varphi$ and rotation of
valence spin by angle $\alpha$, which transform the operators
\begin{equation*}
    \mathcal{C}(\varphi,\alpha): d^{\phantom\dagger}_{ms}\rightarrow 
    e^{im\varphi}e^{i\sigma\alpha}d^{\phantom\dagger}_{m\sigma},
    \quad
    p^{\phantom\dagger}_{m\sigma}\rightarrow 
    e^{i(m+\sigma)\varphi}p^{\phantom\dagger}_{m\sigma},
\end{equation*}
where $m$ is the orbital and $\sigma=\pm\tfrac{1}{2}$ the spin projection 
along the $z$-axis. Next, we use
\begin{equation*}
    \expval{\mathcal{C}(\varphi,\alpha)f|\hT|
    \mathcal{C}(\varphi,\alpha)i}=
    \expval{f|\mathcal{C}^{-1}(\varphi,\alpha)\hT
    \mathcal{C}(\varphi,\alpha)|i}
\end{equation*}
to transform the dipole operators instead of wave functions in (\ref{eq:A}). 
%Note, that for $\varphi=\alpha$ this is a trivial change of the coordinate system, however, for $\varphi\neq\alpha$ we use fact that eigenstates of Hamiltonians with different orientations of the local moment can be transformed into each other, which is possible only 
%in the absence of valence SOC and core-valence multipole interaction. 
Using the definition of the dipole operators
\begin{equation*}
\label{eq:F}
\begin{split}
 \hT^{\pm}&\equiv \hT_\uparrow^{\pm}+ \hT_\downarrow^{\pm}=\sum_{m,\sigma}\Gamma^{\phantom n}_{\pm m} \hd^\dagger_{m\pm 1\sigma}\hp^{\phantom\dagger}_{m\sigma}\\
 \hT^{0}&\equiv\hT_\uparrow^{0}+ \hT_\downarrow^{0}=\sum_{m,\sigma}\Gamma^{(0)}_{m} \hd^\dagger_{m\sigma}\hp^{\phantom\dagger}_{m\sigma},
\end{split}
\end{equation*}
we arrive at their transformation properties
    \begin{equation}
        \begin{split}
     \mathcal{C}^{-1}(\varphi,\alpha)\hT_\sigma^{+}\mathcal{C}(\varphi,\alpha) & =  e^{i\varphi} e^{-i\sigma(\varphi-\alpha)}\hT_\sigma^{+}\\
     \mathcal{C}^{-1}(\varphi,\alpha)\hT_\sigma^{-}\mathcal{C}(\varphi,\alpha) & =  e^{-i\varphi} e^{-i\sigma(\varphi-\alpha)}\hT_\sigma^{-}\\
     \mathcal{C}^{-1}(\varphi,\alpha)\hT_\sigma^{0}\mathcal{C}(\varphi,\alpha) & =  e^{-i\sigma(\varphi-\alpha)}\hT_\sigma^{0}.
     \end{split}
\end{equation}
Substituting these into (\ref{eq:A}) we get the  formula for the XMCD spectra for general angles $\varphi$ and $\alpha$
\begin{equation}
    \begin{split}
    \Delta F(\varphi,\alpha)=&(e^{i\varphi}T_{\uparrow}^+
    -e^{-i\varphi}T_{\uparrow}^-)
    \overline{T_{\uparrow}^0}\\
    &+
    (e^{i\varphi}T_{\downarrow}^+
    -e^{-i\varphi}T_{\downarrow}^-)
    \overline{T_{\downarrow}^0}\\
    &+
    (e^{i\alpha}T_{\uparrow}^+
    -e^{-i(2\varphi-\alpha)}T_{\uparrow}^-)
    \overline{T_{\downarrow}^0}\\
    &+
    (e^{i(2\varphi-\alpha)}T_{\downarrow}^+
    -e^{-i\alpha}T_{\downarrow}^-)
    \overline{T_{\uparrow}^0}+
    c.c.\\
    =&
    (e^{i\alpha}T_{\uparrow}^+
    -e^{-i(2\varphi-\alpha)}T_{\uparrow}^-)
    \overline{T_{\downarrow}^0}\\
    &+
    (e^{i(2\varphi-\alpha)}T_{\downarrow}^+
    -e^{-i\alpha}T_{\downarrow}^-)
    \overline{T_{\uparrow}^0}+
    c.c.
    \end{split}
\end{equation}
The $\uparrow\uparrow$ and $\downarrow\downarrow$ terms do not change sign under the magnetic moment
reversal $\alpha \rightarrow \alpha +\pi$ and thus must vanish the expression for XMCD.

Now we can evaluate the XMCD spectra for the orientations in (\ref{eq:comparsion})
\begin{equation}
\label{eq:directions}
\begin{split}
        & \Delta F(0,0)=
        (T_{\uparrow}^+
    -T_{\uparrow}^-)
    \overline{T_{\downarrow}^0}+
    (T_{\downarrow}^+
    -T_{\downarrow}^-)
    \overline{T_{\uparrow}^0}+
    c.c.\\
    &\Delta F(\tfrac{\pi}{2},\pi)=
            (-T_{\uparrow}^+
    -T_{\uparrow}^-)
    \overline{T_{\downarrow}^0}+
    (T_{\downarrow}^+
    +T_{\downarrow}^-)
    \overline{T_{\uparrow}^0}+
    c.c.\\
    &\Delta F(\tfrac{\pi}{4},\tfrac{\pi}{2})=
            (iT_{\uparrow}^+
    -T_{\uparrow}^-)
    \overline{T_{\downarrow}^0}+
    (T_{\downarrow}^+
    +iT_{\downarrow}^-)
    \overline{T_{\uparrow}^0}+
    c.c.
\end{split}
\end{equation}
Note that this is not enough to guarantee $\Delta F^{(a)}=\Delta F^{(b)}$. We use the fact that
in the rutile structure there are two-fold rotation axes parallel to $[110]$ and $[1\bar{1}0]$, i.e.,
$\varphi=0$ and $\varphi=\tfrac{\pi}{2}$. In Ref.~\onlinecite{Hariki2023} we have shown that such rotation symmetry implies vanishing of XMCD for the magnetic moment perpendicular to the
X-ray propagation vector, i.e., 
%$\Delta F(0,\tfrac{\pi}{2})=0$ and $\Delta F(\tfrac{\pi}{2},\tfrac{\pi}{2})=0$. Using this observation
\begin{align*}
%\begin{split}
        & \Delta F(0,\tfrac{\pi}{2})=\\
        &(iT_{\uparrow}^+
    -iT_{\uparrow}^-)
    \overline{T_{\downarrow}^0}+
    (-iT_{\downarrow}^+
    +iT_{\downarrow}^-)
    \overline{T_{\uparrow}^0}+
    c.c.=0\\
     & \Delta F(\tfrac{\pi}{2},\tfrac{\pi}{2})=\\
        &(iT_{\uparrow}^+
    +iT_{\uparrow}^-)
    \overline{T_{\downarrow}^0}+
    (iT_{\downarrow}^+
    +iT_{\downarrow}^-)
    \overline{T_{\uparrow}^0}+
    c.c.=0. %\\,
%\end{split}
\end{align*}
%This can be rewritten as %we arrive at 
Adding the two lines we get
\begin{equation}
\label{eq:aux}
%\begin{split}
%    &
    iT_{\uparrow}^+
    \overline{T_{\downarrow}^0}
    +iT_{\downarrow}^-
    \overline{T_{\uparrow}^0}+
    c.c.=0,
%    \\
%    &
%    iT_{\uparrow}^-
%    \overline{T_{\downarrow}^0}
%    +iT_{\downarrow}^+
%    \overline{T_{\uparrow}^0}+
%    c.c.=0.
%\end{split}
\end{equation}
and substituting (\ref{eq:aux}) into the third line of (\ref{eq:directions}) concludes the proof
\begin{equation*}
\begin{split}
    \tfrac{1}{2}\left(\Delta F(0,0)+\Delta F(\tfrac{\pi}{2},\pi)\right)&=
    -T_{\uparrow}^-
    \overline{T_{\downarrow}^0}+
    T_{\downarrow}^+
    \overline{T_{\uparrow}^0}
   + c.c.\\
%   &= -2T_{\uparrow}^-
%    \overline{T_{\downarrow}^0}+
%    2T_{\downarrow}^+
%    \overline{T_{\uparrow}^0}\\
     &=\Delta F(\tfrac{\pi}{4},\tfrac{\pi}{2}) %=
%        -2T_{\uparrow}^-
%    \overline{T_{\downarrow}^0}
%    +2T_{\downarrow}^+
%    \overline{T_{\uparrow}^0}.
\end{split}
\end{equation*}
\newline
\bibliography{main}

\end{document}

\title{Supplementary Material to ``X-ray Magnetic Circular Dichroism in \ruo''}

\author{A.~Hariki}
\affiliation{Department of Physics and Electronics, Graduate School of Engineering,
Osaka Metropolitan University, 1-1 Gakuen-cho, Nakaku, Sakai, Osaka 599-8531, Japan}
\author{Y.~Takahashi}
\affiliation{Department of Physics and Electronics, Graduate School of Engineering,
Osaka Metropolitan University, 1-1 Gakuen-cho, Nakaku, Sakai, Osaka 599-8531, Japan}
\author{J.~Kune\v{s}}
\affiliation{Institute for Solid State Physics, TU Wien, 1040 Vienna, Austria}

\maketitle

\section{XMCD simulation for \ruo}

We calculate Ru $M$- and $L$-edge X-ray absorption spectroscopy (XAS) and X-ray magnetic circular dichroism (XMCD) spectra using the LDA+DMFT Anderson impurity model (AIM) and the Ru$^{4+}$ atomic model.
%The absorption of a core-level electron to the valence state is a local excitation, forming a bound exciton between the core hole and Ru 4$d$ electrons on the X-ray excited site, and the core-hole on the  Ru 2$p$ or 3$p$ level does not leave the excited Ru site. Thus,  single impurity model, even the atomic model, provides its good description~\cite{Groot90,groot_kotani,Winder20}.
%Indeed it was demonstrated in early study that the Ru $L$-edge XAS spectra of RuO$_2$ and some Ru compounds can be well fitted within the atomic model~\cite{Hu00}. 
%To complement the atomic-model simulation and demonstrate that the hybridization effect is not essential for the XMCD analysis, we present the LDA+DMFT AIM simulation which incorporates the hybridization of the Ru 4$d$ states on the X-ray excited site with the spin-polarized Ru bands in the antiferromagnetic long-range order.
Firstly, the LDA bands obtained with the WIEN2K package~\cite{wien2k} for the experimental rutile structure~\cite{Berlijn17} are projected onto a tight-binding model spanning Ru 4$d$ bands, encompassing both the Ru $t_{2g}$ and $e_g$ states~\cite{wannier90,wien2wannier}.
The multi-band Hubbard model is defined by incorporating local electron-electron interaction within the Ru 4$d$ shell to the tight-binding Hamiltonian. The electron-electron interaction is parameterized by Hubbard $U=F_0$ and Hund’s $J=(F_2+F_4)/14$ parameters where $F_0$, $F_2$, and $F_4$ are the Slater integrals~\cite{Pavarini1,Pavarini2}.
We adopt $U=3.0$~eV and $J=0.45$~eV which yields an antiferromagnetic solution with a magnetic moment consistent with Ref.~\cite{Ahn19}. %\textcolor{red}{A cure for the large magnetic moment?} 
The chosen $U$ value is slightly larger than that in Ref.~\cite{Ahn19} due to explicit treatment of the Ru 4$d$ $e_g$ states in the multi-band Hubbard model explicitly. 
While the inclusion of Ru 4$d$ $e_g$ states is not essential for describing the magnetic and electronic structure of the system, it proves crucial for the XMCD line shape. This is because an X-ray excited core electron can be absorbed into the Ru $e_g$ orbitals, not just the $t_{2g}$ orbitals. 
%The atomic models for the Ru1 and Ru2 sites implement the onsite term of the multi-band Hubbard model.  
A standard LDA+DMFT calculation~\cite{Kotliar06,Georges96} for the multi-band Hubbard model is performed, following the same implementation of Refs.~\cite{Hariki2017,Winder20}.
The continuous-time quantum Monte Carlo method with the hybridization expansion formalism ~\cite{werner06,boehnke11,hafermann12} is employed to solve the auxiliary AIM in the DMFT self-consistent calculation. 
The valence spectral intensities and hybridization densities $\Delta (\omega)$ are computed on the real frequency axis after the self-energy is analytically continued using the maximum entropy method~\cite{wang09,jarrell96}.
Subsequently, the XMCD intensities are computed from the AIM with the LDA+DMFT hybridization densities $\Delta (\omega)$, where the Ru 2$p$ ($L$-edge) or 3$p$ ($M$-edge) core orbitals and core-valence interaction are included. 
The core-valence interaction and SOC parameters are estimated by an atomic Hartree-Fock calculation following Refs.~\cite{Hariki2017,Hariki20,Winder20}.
%%%%%%%%%%%%%%%%%%%%%%%%%%%%%%%%%%%%
%%%%%%%%%%%%%%%%%%%%%%%%%%%%%%%%%%%%
\begin{figure}[b]
\includegraphics[width=0.90\columnwidth]{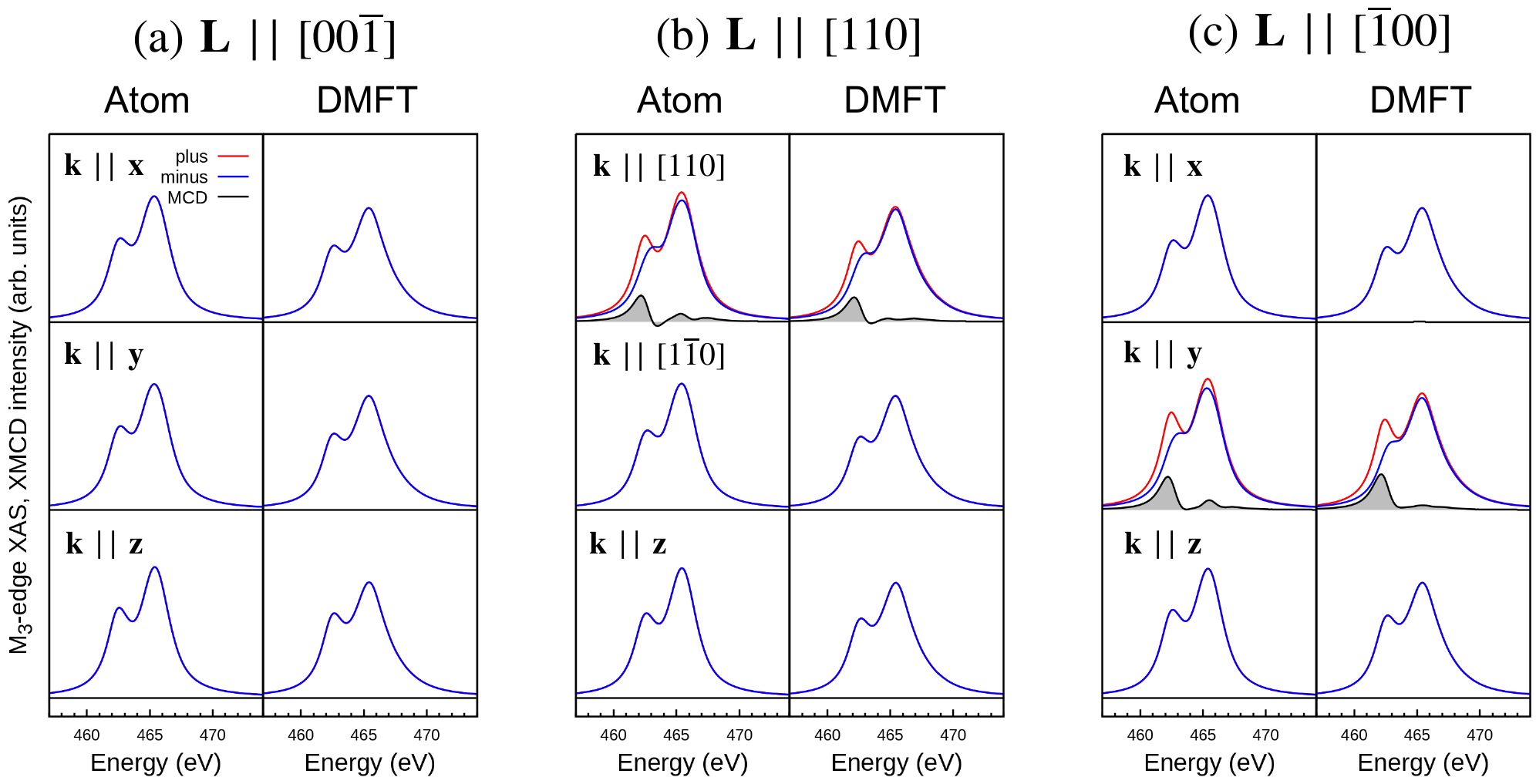}
    \caption{Ru $M_3$-edge XMCD spectra calculated by the Ru atomic model and the LDA+DMFT AIM model for for different orientation of the N\'eel vectors $\bL$ and x-ray wave vectors $\bf k$. }
\label{fig_aim_atom}
\end{figure}
%%%%%%%%%%%%%%%%%%%%%%%%%%%%%%%%%%%%
%%%%%%%%%%%%%%%%%%%%%%%%%%%%%%%%%%%%
The XAS spectral function is described by the Fermi's golden rule 
%%%%%%%%%%%%%%%%%%%%%%%%%%%%%%%%%%%%%%%%%%%%%%%%%
\begin{equation}
\label{eq:xas}
\begin{split}
F^{(i)}_{\rm XAS}(\omega_{\rm in})&=-\frac{1}{\pi}\operatorname{Im} \langle g | T^{\dag} \frac{1}{\omega_{\rm in}+E_g-H_{\rm imp}^{(i)}} T| g \rangle. \notag \\
\end{split}
\end{equation}
%%%%%%%%%%%%%%%%%%%%%%%%%%%%%%%%%%%%%%%%%%%%%%%%
Here, $|g\rangle$ is the ground with energy $E_g$, and $\omega_{\rm in}$ is the energy of the incident photon. 
$T$ is the electric dipole transition operator for circularly-polarized X-rays~\cite{Matsubara00,groot_kotani}. 
The index $i$ ($=1,2$) denotes the X-ray excited Ru site in the unit cell. 
The impurity Hamiltonian $H_{\rm imp}^{(i)}$ for each Ru site is used to calculate the individual site contribution, and the total XMCD intensities are obtained by summing up the contributions of two sites~\cite{Winder20,Hariki2023}.

%%%%%%%%%%%%%%%%%%%%%%%%%%%%%%%%%%%%
%%%%%%%%%%%%%%%%%%%%%%%%%%%%%%%%%%%%
\begin{figure}[t]
\includegraphics[width=0.80\columnwidth]{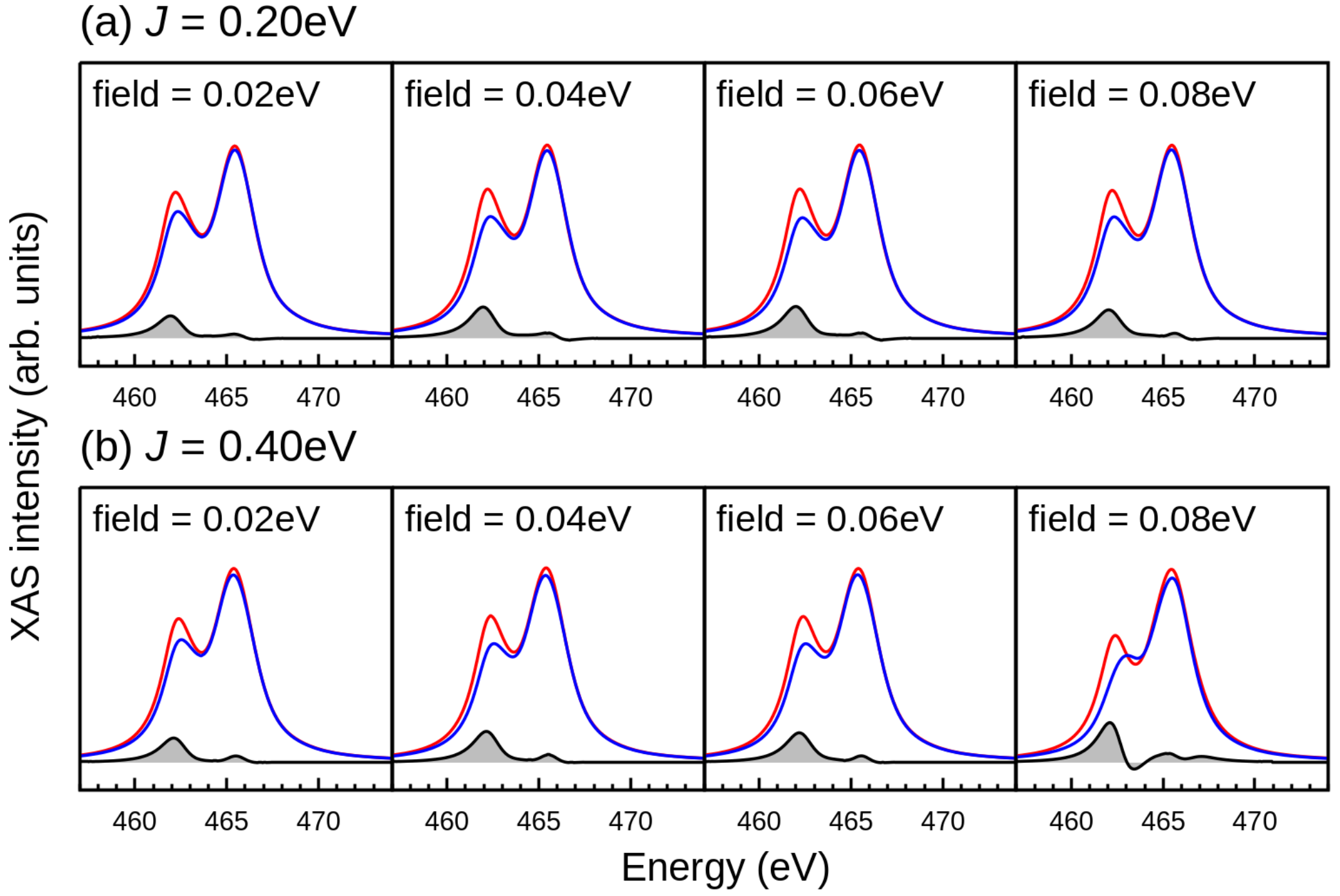}
    \caption{Hund's $J$ parameter and magnetic-field amplitude dependence of the $M_3$-edge XMCD intensities calculated by the Ru atomic model. The calculations are performed for the N\'eel vector $\bL \ || \ [110]$ and x-ray wave vector ${\bf k} \ || \ [110]$. }
\label{fig_hund}
\end{figure}
%%%%%%%%%%%%%%%%%%%%%%%%%%%%%%%%%%%%
%%%%%%%%%%%%%%%%%%%%%%%%%%%%%%%%%%%%

Figure~\ref{fig_aim_atom} shows the calcualted $M_3$-edge XMCD intensities by the Ru$^{4+}$ atomic model and the LDA+DMFT AIM model. The results from the atomic model closely resemble the LDA+DMFT AIM ones for all the studied N\'eel vectors $\bL$ and x-ray wave vectors ${\bf k}$. The amplitude of the Zeeman field and the Hund's coupling $J$ in the atomic model is tuned to give rise to similar XMCD intensities of the LDA+DMFT AIM. Figure~\ref{fig_aim_atom} demonstrates the validity of the use of a computationally-cheaper atomic model to analyze the optical conductivity tensors in the main text. 
Figure~\ref{fig_hund} presents the Hund's $J$ and Zeeman field amplitude dependence of the XCMD intensities in the atomic model.
%Figure~\ref{fig_spin} summarized the values of the $S_x$, $S_y$ and $S_z$ components at the two Ru sites for the studied N\'eel vectors $\bL$ obtaiend with the different amplitudes of the field in the atomic model. 

\begin{comment}

%%%%%%%%%%%%%%%%%%%%%%%%%%%%%%%%%%%%
%%%%%%%%%%%%%%%%%%%%%%%%%%%%%%%%%%%%
\begin{figure}[h]
\includegraphics[width=0.85\columnwidth]{RuO2_figures/fig_spin.pdf}
    \caption{spin expectation()}
\label{fig_spin}
\end{figure}
%%%%%%%%%%%%%%%%%%%%%%%%%%%%%%%%%%%%
%%%%%%%%%%%%%%%%%%%%%%%%%%%%%%%%%%%%
\begin{figure}[h]
\includegraphics[width=0.85\columnwidth]{RuO2_figures/fig_spin_no_SOC.pdf}
    \caption{spin expectation(no SOC)}
\label{fig_spin_no_SOC}
\end{figure}
%%%%%%%%%%%%%%%%%%%%%%%%%%%%%%%%%%%%
%%%%%%%%%%%%%%%%%%%%%%%%%%%%%%%%%%%%

\end{comment}

%\begin{figure}[h]
%\includegraphics[width=0.45\columnwidth]{RuO2_figures/fig_interaction.pdf}
%    \caption{Ru }
%\label{fig_interaction}
%\end{figure}
%%%%%%%%%%%%%%%%%%%%%%%%%%%%%%%%%%%%
%%%%%%%%%%%%%%%%%%%%%%%%%%%%%%%%%%%%
%\begin{figure}[h]
%\includegraphics[width=0.85\columnwidth]{RuO2_figures/fig_interaction2.pdf}
%    \caption{Ru }
%\label{fig_interaction}
%\end{figure}
%%%%%%%%%%%%%%%%%%%%%%%%%%%%%%%%%%%%
%%%%%%%%%%%%%%%%%%%%%%%%%%%%%%%%%%%%
%\begin{figure}[h]
%\includegraphics[width=0.85\columnwidth]{RuO2_figures/fig_interaction_Mn.pdf}
%    \caption{Ru }
%\label{fig_interaction}
%\end{figure}
%%%%%%%%%%%%%%%%%%%%%%%%%%%%%%%%%%%%

\bibliography{main}